\documentstyle[12pt]{article}
\def\lulu{{\tt SABSPV} }
\begin{document}
\begin{titlepage}
\rightline{CERN-TH/95-169}
\vspace{0.5cm}
\noindent
{\Large{\bf \lulu $-$ A Monte Carlo integrator \\
for Small-Angle Bhabha Scattering}
}
\vspace{1cm}
\bigskip\bigskip

\noindent
Matteo CACCIARI~\footnote{\footnotesize
Della Riccia and Universit\`a di Pavia Fellow. Work supported by INFN, Sezione
di Pavia, Italy.} \\
{\it DESY - Deutsches Elektronen-Synchrotron } \\
{\it Notkestrasse 85 - 22607 HAMBURG - GERMANY} \\
\medskip\medskip

\noindent
Guido MONTAGNA \\
{\it Dipartimento di Fisica Nucleare e Teorica, Universit\`a di Pavia } \\
{\it Via A.~Bassi n.~6 - 27100 PAVIA - ITALY} \\
\medskip\medskip

\noindent
Oreste NICROSINI~\footnote{\footnotesize
On leave from INFN, Sezione di Pavia, Italy.} \\
{\it CERN, Theory Division} \\
{\it CH 1211 - Geneva 23 - SWITZERLAND} \\
\medskip\medskip

\noindent
Fulvio PICCININI  \\
{\it Istituto Nazionale di Fisica Nucleare, Sezione di Pavia } \\
{\it Via A.~Bassi n.~6 - 27100 PAVIA - ITALY} \\

\noindent
Program classification: 11.1 \\
\bigskip\bigskip

\noindent
{\small
\lulu  \ is a code designed to perform a theoretical evaluation of
small-angle Bhabha scattering cross sections by suitably matching fixed-order
perturbative
calculations and structure-function techniques.
The implementation of realistic experimental triggering conditions is
achieved by using Monte Carlo integration.
}
\medskip\medskip

\begin{center}
Submitted to Computer Physics Communications
\end{center}

\vfil
\leftline{CERN-TH/95-169}
\leftline{July 1995}
\end{titlepage}

\vfil
\eject

\leftline{\Large{\bf PROGRAM SUMMARY}}
\vskip 15pt

\leftline{{\it Title of program:} \lulu}
\vskip 8pt

\noindent
{\it Computer:} DEC VAX, HP/APOLLO, IBM/RS6000;
{\it Installation:} INFN, Sezione di Pavia, via A.~Bassi 6, 27100 Pavia, Italy
\vskip 8pt

\leftline{{\it Operating system:} VMS, UNIX}
\vskip 8pt

\leftline{{\it Programming language used:} FORTRAN 77}
\vskip 8pt

\noindent
\leftline{{\it Memory required to execute with typical data:} 40 kByte}
\noindent

\vskip 8pt

\leftline{{\it No. of bits in a word: } 32}
\vskip 8pt

\leftline{\it The code has not been vectorized}
\vskip 8pt

\noindent
{\it Subprograms used:} random number generator {\tt RANLUX}~[1];
routines from the CERN Program Library are also used.
\vskip 8pt

\leftline{{\it No. of lines in distributed program, including test data
etc.:} 1700. }
\vskip 8pt

\leftline{\it Correspondence to:}
\leftline{CACCIARI@DESY.DE, CACCIARI@PV.INFN.IT;}
\leftline{MONTAGNA@PV.INFN.IT;}
\leftline{NICROSINI@VXCERN.CERN.CH, NICROSINI@PV.INFN.IT;}
\leftline{PICCININI@PV.INFN.IT}
\vskip 8pt

\noindent
{\it Keywords:} high energy electron--positron collisions,
small-angle Bhabha scattering,
luminosity determination,
 QED corrections, electron structure functions,
 experimental cuts, Monte Carlo integration.

\vskip 8pt

\leftline{{\it Nature of physical problem} }
\noindent
The precise determination of the theoretical Bhabha scattering cross section
in the small-angle regime is a key ingredient for precision luminometry at
LEP. To this aim, QED radiative corrections to the tree-level Standard Model
Cross section have to be taken into account, together with vacuum
polarization effects. Particular care has to be devoted to higher-order
corrections. The theoretical formulation must allow the computation of the
cross section for a wide variety of cuts on the final-state particles.

\noindent
\vskip 8pt
\leftline{{\it Method of solution} }
\noindent
A suitable matching of a fixed-order
calculation with the structure-function techniques for resumming large
initial- and final-state
leading logarithmic corrections~[2] is performed. A Monte Carlo integration
with weighted events has been implemented in order to mimic as close as
possible the experimental triggering conditions.  The importance-sampling
technique~[3]
is employed to take care of the peaking behaviour of the integrand.

\vskip 8pt
\leftline{{\it Restrictions on the complexity of the problem} }
\noindent
The $\cal O (\alpha)$ QED corrections are computed exactly only for the
dominant contribution to the small-angle cross section, namely the square of
$\gamma$-exchange in the $t$-channel; all the other contributions to the
cross sections are corrected at the leading-logarithmic level. The
contribution of additional hadronic or leptonic pairs is at present
neglected. Starting from ${\cal O} (\alpha^2)$, QED corrections are implemented
at the leading-logarithmic level.

\vfil \eject
\leftline{\it Typical running time}
\noindent
On a HP 9000/735 the code takes about $2\times 10^{-4}$ seconds per event, with
standard cuts. A
one per mille relative error can be achieved with about $10^7$ events.
\noindent

\vskip 8pt

\leftline{{\it Unusual features of the program:} none}
\noindent
\vskip 8pt

\leftline{{\it References} }
\noindent
\vskip 8pt \noindent
[1] F.~James, Comput. Phys. Commun. 79 (1994) 111.

\vskip 10pt \noindent
[2] M.~Cacciari, G.~Montagna, O.~Nicrosini and F.~Piccinini, in ``Reports of
the Working Group on Precision Calculations for the $Z$ Resonance'', eds.
D.~Bardin, W.~Hollik and G.~Passarino (CERN  Report 95-03, Geneva, 1995),
p.~389, and references therein.

\vskip  10pt \noindent
[3] F.~James, Rep. Prog. Phys. 34 (1980) 1145.
\vfil
\eject

\leftline{\Large{\bf LONG WRITE -- UP}}
\vskip 15pt

\section{Introduction}
\vskip 10pt

Precision luminometry at LEP requires the best possible knowledge of the
theoretical Bhabha scattering cross section in the small-angle regime.
Actually, the experimental precision of luminosity measurements is at present
at the level of 0.1\% and is foreseen to improve further in the near future.
In order to exploit such an experimental precision, theoretical results of
comparable accuracy are required.

Two kinds of problems are to be faced here. The first one is to establish the
physical precision of a given theoretical approach; this requires an estimate
of those contributions to the cross section which are neglected by the
approach itself. The second one is to establish the technical precision of
the computational realization of a given theoretical approach; this requires
an estimate of the errors introduced by numerical approximations, different
implementations of theoretically known effects and so on, and, last but not
least, bugs in the computer code.

The comparison between different and independent approaches can contribute
to the solution of both problems and become important in order, on the one
hand, to assess the presently reached accuracy and, on the other, to
eventually improve it in a significant way.

Some codes that compute the small-angle Bhabha scattering cross section
already exist~\cite{bhlumi,bhagen}. The aim of the present work is to
describe in some detail the code \lulu.

In order to compute a small-angle Bhabha scattering cross section with a
precision of the order of 0.1\% and within realistic experimental cuts,
the following tasks have to be accomplished:

\begin{enumerate}
\item to use a theoretical formulation capable of taking into account the
effects of multiple soft and/or
collinear photon emission from initial- and final-state particles,
but also of providing an exact evaluation of at least one hard photon emission;
\item to perform an implementation of the integration within the phase-space
limits such that a wide variety of experimental triggering conditions can be
closely reproduced.
\end{enumerate}

The first point is solved by developing a formulation, described in~\cite{yr},
which allows the matching of a fixed-order perturbative calculation with the
structure-function technique. The outcome of this matching is a calculation
which correctly describes one hard photon emission but also allows for the
evaluation of multiphoton emission in the initial and final state.

As for the second point, the cuts are implemented by using a Monte Carlo
integration. The four-momenta of the outgoing particles (electron, positron and
photon) are constructed and fed to a cutting routine that reproduces the
experimental triggering conditions.

In the following we will first review the theoretical background and its
implementation in the code. The structure of the integrand demands the
importance-sampling technique to be applied in order to have an efficient
Monte Carlo integration. We will discuss the way this has been done. Finally,
the
program structure will be briefly presented and the  required inputs and the
corresponding output will be described.

\section{Theoretical Background}
\vskip 10pt

The structure-function approach allows the leading-log-corrected cross
section in the laboratory frame to be written,
taking into account all-orders
photonic radiation, in the following form:
\begin{eqnarray}
\sigma^{(\infty)}_{LL}  &=& \int d \Omega
d x_1 d x_2
D^{(\infty)} (x_1,Q^2) D^{(\infty)} (x_2,Q^2) J(x_1,x_2,\vartheta) \nonumber \\
&\times& {{d \sigma} \over {d \Omega}}
( \hat s (x_1,x_2), \hat t (x_1,x_2,\vartheta) ) F^{(\infty)} (\xi_1, Q^2)
F^{(\infty)} (\xi_2, Q^2) \Theta ({\hbox{\rm cuts}}) .
\label{eq:master}
\end{eqnarray}
The detailed description of Eq.~(\ref{eq:master}) can be found
in~\cite{yr,prd}.  For convenience, we recall that $\sigma$ is the Born
cross section, the $D$'s are the structure functions for initial-state
radiation
and the $F$'s describe final-state photonic emission;
$\Theta ({\hbox{\rm cuts}})$
represents the rejection algorithm for
implementing the experimental cuts.

It was prompted in the Introduction that when aiming at a per mille accuracy we
have to rely also on a complete $\cal O(\alpha)$ calculation of small-angle
Bhabha
scattering cross section. The complete squared matrix element
of the (gauge-invariant) $t$-channel diagrams associated
with (hard) photon radiation from the electron line
has been computed in~\cite{noi}.
Its full expression, including all
${\cal O}(m^2)$, ${\cal O}(m^4)$ and ${\cal O}(m^6)$ terms,
reads as follows:
\begin{eqnarray}
\vert M \vert^2 = {{2 (4 \pi \alpha)^3} \over {t^2}} \Bigg\{ M_0 + m^2 M_2
+ m^4 M_4 + m^6 M_6 \Bigg\} \; , \label{eq:me}
\end{eqnarray}
where the single terms $M_i$ are explicitly reported in~\cite{noi}.
{}From this matrix element the $\cal O(\alpha)$ cross section can be computed
via Monte Carlo integration as
\begin{equation}
\sigma_\gamma^H ( k_0)
= \int d [PS] \> F \Big( \vert M_- \vert^2
+ \vert M_+ \vert^2 \Big)  \Theta ({\hbox{\rm cuts}}) \; ,
\label{eq:bharad}
\end{equation}
where $d [PS]$ is the phase-space volume element,
$\vert M_\pm \vert^2$ are the squared amplitudes for
the electronic and positronic radiation, $F$ is the
proper flux factor, and $k_0$ is a minimum energy fraction of
the radiated photon, defined as $k_0 = E^\gamma_{min} / E$.

The matching between the all-orders leading-log cross section of
eq.~(\ref{eq:master}) and the $\cal O(\alpha)$ one (\ref{eq:bharad})
takes place in
the following way: the order-$\alpha$ content of the leading-log
$t$-channel $\gamma$-exchange cross section
is extracted by employing the $\cal O(\alpha)$ expansions of the structure
functions
entering the master formula (\ref{eq:master}), thereby yielding
$\sigma^{(\alpha)}_{LL,\,\gamma}$. Denoting by $\sigma^{S+V}_\gamma (k_0)$
the  $t$-channel $\gamma$-exchange cross section
 including virtual corrections plus soft photons of energy up to
$E_\gamma = k_0 E$ (see for instance~\cite{cr}),
the cross section in the luminometry region can finally be written as
\begin{equation}
\sigma_A  = \sigma^{(\infty)}_{LL}
 - \sigma^{(\alpha)}_{LL,\,\gamma} +
\sigma^{S+V}_\gamma (k_0) + \sigma^{H}_\gamma (k_0) \; ,
\label{eq:add}
\end{equation}
where, we recall,  $\sigma^{(\infty)}_{LL}$ is the all-orders
leading-log-corrected
full Bhabha cross section of Eq.~(\ref{eq:master}),
$ \sigma^{(\alpha)}_{LL,\,\gamma}$ is
the up to $\cal O (\alpha)$ leading-log-corrected cross section, limited to
the $t$-channel $\gamma$-exchange contribution, and $ \sigma^{H}_\gamma (k_0)$
is the radiative Bhabha cross section of
Eq.~(\ref{eq:bharad}).

The difference $\sigma^{(\infty)}_{LL} -  \sigma^{(\alpha)}_{LL,\,\gamma}$
contains the Born contribution of all the Bhabha
channels but the $t$-channel $\gamma$-exchange one, plus their leading-log QED
corrections resummed to all orders, plus the higher-order leading-log QED
corrections to  the $t$-channel $\gamma$-exchange contribution starting from
${\cal O} (\alpha^2)$; the exact up to $\cal O ( \alpha)$ contribution for
  the $t$-channel $\gamma$-exchange term is then supplied by
$\sigma^{S+V}_\gamma (k_0) + \sigma^{H}_\gamma (k_0)$.

Equation~(\ref{eq:add}) is in the additive form. A factorized form can also be
supplied. It has the same $\cal O ( \alpha)$ content but also leads to
the so-called classical limit, according to which the cross section
must vanish in the absence of photonic radiation. It reads
\begin{eqnarray}
&&\sigma_F  = ( 1 + C_{NL}^H ) \sigma^{(\infty)}_{LL} , \nonumber \\
&&\nonumber \\
&&C_{NL}^H \equiv {{\sigma^{S+V}_\gamma (k_0) + \sigma^{H}_\gamma (k_0)
- \sigma^{(\alpha)}_{LL,\,\gamma} }\over \sigma} \equiv
{{\sigma_{NL,\gamma}^{(\alpha)}}\over{\sigma}} ,
\label{eq:fact}
\end{eqnarray}
$\sigma$ being the Born cross section; $C_{NL}^H$ contains the non-log part of
the $\cal O ( \alpha)$ $\gamma(t)\gamma(t)$ cross section, represented by
$\sigma_{NL,\gamma}^{(\alpha)}$. A detailed discussion on the physical
content of eqs.~(\ref{eq:add}) and~(\ref{eq:fact}) will be given elsewhere.

\section{Importance Sampling}
\vskip 10pt

Both the leading-log cross section and the real hard photon one present
singularities which have to be properly treated with importance-sampling
techniques to ensure a reasonably fast convergence of the Monte Carlo
integration. This means dividing the integrand by a so-called weight
function and redefining the integration variable:
\begin{equation}
\int f(x)dx = \int{{f(x)}\over{p(x)}} p(x)dx \to \int{{f(y)}\over{p(y)}} dy
\end{equation}
with $y = \int p(x)dx$. The weight function $p(x)$ must be chosen such that the
ratio $f(x)/p(x)$ is smooth enough to
produce a small variance.

Let us start with the leading-log cross section. The structure function
$D^{(\infty)} (x,Q^2)$ reads
\begin{eqnarray}
D^{(\infty)} (x,Q^2) &=& {{\exp \Big\{ {1 \over 2} \beta ( {3 \over 4}
- \gamma_E )
\Big\}} \over {\Gamma \big( 1 + {1 \over 2} \beta \big) }}
{{1} \over {2}} \beta
(1-x)^{{{1} \over {2}} \beta -1}
- {{1} \over {4}} \beta (1+x) \nonumber \\
& & + {{1} \over {32}}
\beta^2 \Big[ -4 (1+x) \ln (1-x)
+  3 (1+x) \ln x \Big. \nonumber \\
& & - \Big. 4 {{\ln x} \over {1-x}} - 5 - x \Big] \; ,\qquad\qquad
\beta = {{2\alpha}\over{\pi}}
\left[\ln\left({{Q^2}\over{m_e^2}}\right)-1\right]\;,
\label{eq:strucfun}
\end{eqnarray}
and displays a singularity in $x=1$. This is disposed of by
employing, in connection with each structure function in the integrand,
a weight function that resembles the singular  part of the structure
function itself:
\begin{equation}
p(x) = {{1} \over {2}} \beta
(1-x)^{{{1} \over {2}} \beta -1}   .
\end{equation}
The relation between the original integration variable $x$ and the flat random
variable $r\in(0,1)$ is then
\begin{equation}
x = 1 -
(1-r)^{2/\beta} .
\end{equation}

The real hard photon cross section evaluation must instead face
the ``singular''(in fact, only extremely peaked)
distribution of the relative angle $\theta_\gamma$ between the photon and
the emitting particle when approaching the forward collinear region. Four such
``singularities'' are present, related to initial- and
final-state emission from
electron and positron respectively. The initial-state ones are smoothed by
choosing the weight function
\begin{equation}
p(\cos\theta_\gamma) = \Big(1-\sqrt{1-m_e^2/E^2}\cos\theta_\gamma\Big)^{-1}  ,
\end{equation}
$m_e$ and $E$ being the electron mass and energy in the lab frame,
respectively.
The final-state ones are disposed of by using
similar weight functions after a proper rotation of the reference frame such
that the $z$-axis is taken along the outgoing lepton.

\section{Program Structure}
\vskip 10pt

The code has the usual structure of Monte Carlo programs, with
routines evaluating the cross sections (some pertaining
to the real hard photon part, some to the leading-log cross section),
triggering routines,
determining whether the event is to be accepted or not, utility
routines, and finally routines
collecting the output. More in detail, we have:
\begin{itemize}
\item Monte Carlo routines:
\begin{itemize}
\item the {\tt MAIN PROGRAM}: where the Monte Carlo loop is located.
It provides the random numbers to the routine {\tt INTEGR}, collects its
outputs, performs the
statistics operations and writes the results to the output file;
\item {\tt SUBROUTINE INTEGR}: receives the random numbers from the {\tt MAIN
PROGRAM} and calls the cross sections, kinematics and triggering routines. It
returns the cross sections for a given event to the {\tt MAIN PROGRAM}.
\end{itemize}
\item Cross section routines:
\begin{itemize}
\item {\tt SUBROUTINE HARD}, {\tt SUBROUTINE EEG}: provide the real hard photon
cross section;
\item {\tt FUNCTION BHABORN}, {\tt FUNCTION BHABORN\_gtgt}, {\tt FUNCTION
AINTSV}: provide the full Born cross section, the Born cross section for the
$\gamma(t)\gamma(t)$ channel and the virtual plus soft $\cal O(\alpha)$
correction
to the latter, respectively;
\item {\tt FUNCTION DOP}, {\tt FUNCTION AJAC}, {\tt FUNCTION PDINFTY}, {\tt
FUNCTION PDPDALPHA}: routines related to the structure-function evaluation of
the leading-log cross section;
\item {\tt FUNCTION VPOL}, {\tt FUNCTION SOMMAT}, {\tt FUNCTION ABC}: evaluate
the vacuum polarization correction to the photon propagator, according to the
parametrization of~\cite{vacpol}.
\end{itemize}
\item Kinematic routines:
\begin{itemize}
\item {\tt SUBROUTINE KINE2}, {\tt SUBROUTINE QUADRI2}: build up the outgoing
particles four-momenta from the generated random numbers for two-body events,
i.e. Born and virtual plus soft cross sections, and collinear leading-log
ones;
\item {\tt SUBROUTINE KINEH}, {\tt SUBROUTINE QUADRIH}: the same but for
three-body events, with the real hard photon.
\end{itemize}
\vfil \eject
\item Triggering routines:
\begin{itemize}
\item {\tt SUBROUTINE CUTS}: master cut routine. It propagates to the rest of
the code the result of one of the triggering routines;
\item {\tt SUBROUTINE TRIGGER1}: performs acceptance cuts for a sample
trigger;
\item {\tt SUBROUTINE TRIGGER2}: dummy routine to be used to implement a new
trigger.
\end{itemize}
\item Utility routines:
\begin{itemize}
\item{\tt FUNCTION EXCHANGE}, {\tt FUNCTION COSANGLE}, {\tt FUNCTION PHIROT}:
they exchange any two variables, evaluate the cosine of the angle between
any two four-momenta and rotate a four-momentum around its $z$-axis,
respectively.
\end{itemize}
\end{itemize}

The code can be used in a ``black box'' way with the only exception of the
triggering routine, which may be provided by the user. An example of how to
build such a routine is given in Section~\ref{trigger}. We now want to
describe the
common blocks the user has to be familiar with for writing his own triggering
routine.
\begin{itemize}
\item
\begin{verbatim}
      REAL*8 QUADRIM(0:3), QUADRIP(0:3), KUADRI(0:3)
      COMMON/MOMENTA/QUADRIM,QUADRIP,KUADRI
\end{verbatim}
This common block contains the four-momenta of the outgoing electron ({\tt
QUADRIM}), positron ({\tt QUADRIP}) and photon ({\tt KUADRI}). It is filled by
the {\tt QUADRI2} or {\tt QUADRIH} routines when treating
soft-collinear or real hard photons respectively.

\item
\begin{verbatim}
      COMMON/CUTOFF/EGMIN,IHARD
\end{verbatim}
These are non-physical parameters. {\tt EGMIN} contains
the value of the cutoff $k_0$, which separates virtual
plus soft and real hard photon events. {\tt
IHARD} is a flag which is set to {\tt 1} when processing real hard photon
events, and {\tt 0} otherwise. It can be used in the
triggering routines to skip
or implement some cuts according to the event
being of two-body/collinear type or with a
real hard photon. For example, the cut on the $k_0$ cutoff only has to be
activated in the real hard photon part, being  taken into account
analytically in the virtual plus soft part.

\item
\begin{verbatim}
      COMMON/CONST/EBEAM,ALPHA,CONVFAC,PI,AME
\end{verbatim}
Some constants, which are used in many points of the code. {\tt EBEAM} is the
beam energy in GeV (see Section~\ref{input}
about input parameters). {\tt ALPHA} is the
electromagnetic coupling, $\alpha = 1/137.0359895$. {\tt CONVFAC} is the
conversion factor from GeV$^{-2}$ to nb: $.38937966\times 10^{6}$. {\tt
Pi} is $\pi$ and, finally, {\tt AME} is the electron mass: $m_e =
0.51099906\times 10^{-3}$~GeV.

\item
\begin{verbatim}
      COMMON/EXPCUTS/T1MIN,T1MAX,T2MIN,T2MAX,E1MIN,E2MIN
\end{verbatim}
The experimental acceptance cuts. See Section~\ref{input} about inputs
description.

\item
\begin{verbatim}
      REAL*8 CALOINPUT(5)
      COMMON/CALOS/CALOINPUT
\end{verbatim}
These are parameters which are read in from the input file (once more, see
Section~\ref{input}) and which can be used in writing the triggering routines.
The meaning
of the five parameters will depend on the use which is made of them in the
routines. For instance, in {\tt TRIGGER1} the parameter {\tt CALOINPUT(1)} is
the
calorimetric threshold as a fraction of the beam energy and
{\tt CALOINPUT(2)} is the half-opening (in radians) of the cone,
which defines an electromagnetic cluster.
\end{itemize}

\section{Sample Trigger Routine}
\label{trigger}
\vskip 10pt

A sample triggering routine {\tt TRIGGER1()} is inserted in the code. It can be
used as an example to construct other experimental-like triggers. A dummy
{\tt TRIGGER2()} is also given, to be completed by the user.

Below we give the listing of {\tt TRIGGER1()}. It first checks for the
electron and the positron to be within angular acceptance cuts. After that, it
checks whether one of the leptons and the photon do form an electromagnetic
cluster, i.e. are radially separated by less than
the critical distance fixed by
the input parameter {\tt DEL}. If they do, the energy  cut is
performed using the sum of their energies rather than the energy of the bare
fermion only.

\begin{verbatim}
      LOGICAL FUNCTION TRIGGER1()
      IMPLICIT REAL*8 (A-H,O-Z)
      REAL*8 QUADRIM(0:3), QUADRIP(0:3), KUADRI(0:3)
      REAL*8 ZETA(0:3)
      REAL*8 CALOINPUT(5)
      COMMON/MOMENTA/QUADRIM,QUADRIP,KUADRI
      COMMON/EXPCUTS/T1MIN,T1MAX,T2MIN,T2MAX,E1MIN,E2MIN
      COMMON/CUTOFF/EGMIN,IHARD
      COMMON/CONST/EBEAM,ALPHA,CONVFAC,PI,AME
      COMMON/CALOS/CALOINPUT
      DATA ZETA/0D0,0D0,0D0,1D0/
*
      EGAMMA = KUADRI(0)
*
      IF(IHARD.EQ.1) THEN
*
*.....THIS IS THE NON-PHYSICAL CUTOFF
*
        IF(EGAMMA.LT.EGMIN) THEN
           TRIGGER1 = .FALSE.
           RETURN
        ENDIF
      ENDIF
*
      THETA1 = ACOS(COSANGLE(QUADRIM,ZETA))
      THETA2 = ACOS(COSANGLE(QUADRIP,ZETA))
      THETA2OPP = PI-THETA2
      E1 = QUADRIM(0)
      E2 = QUADRIP(0)
*
      IF(THETA1.LT.T1MIN) THEN
         TRIGGER1 = .FALSE.
         RETURN
      ENDIF
*
      IF(THETA1.GT.T1MAX) THEN
         TRIGGER1 = .FALSE.
         RETURN
      ENDIF
*
      IF(THETA2OPP.LT.T2MIN) THEN
         TRIGGER1 = .FALSE.
         RETURN
      ENDIF
*
      IF(THETA2OPP.GT.T2MAX) THEN
         TRIGGER1 = .FALSE.
         RETURN
      ENDIF
*
      EMCL = E1
      EPCL = E2
*
      IF (IHARD.EQ.1) THEN
        DJET = CALOINPUT(2)
        CDJET = COS(DJET)
        CEG = COSANGLE(QUADRIM, KUADRI)
        CPG = COSANGLE(QUADRIP, KUADRI)
*
        IF(CEG.GE.CDJET) THEN
           EMCL = E1 + EGAMMA
        ENDIF
*
        IF(CPG.GE.CDJET) THEN
           EPCL = E2 + EGAMMA
        ENDIF
*
      ENDIF
*
      CALOTH = CALOINPUT(1)
      SUMEN = EMCL*EPCL/EBEAM/EBEAM
*
      IF(SUMEN.LT.CALOTH) THEN
         TRIGGER1 = .FALSE.
         RETURN
      ENDIF
*
      TRIGGER1 = .TRUE.
*
      END
\end{verbatim}

\section{Input Description}
\label{input}
\vskip 10pt

A data card of the following kind has to be provided, via standard input,
when running the program:
\begin{verbatim}
46.15D0                                 ! EBEAM
24.D-3  58.D-3  0.D0                    ! T1MIN, T1MAX, E1MIN
24.D-3  58.D-3  0.D0                    ! T2MIN, T2MAX, E2MIN
0.5D0   1.D-2   0.D0  0.D0  0.D0        ! CALOINPUT(1...5)
1                                       ! ISIM
1                                       ! ICALO
1.D5    0.D0    0     'SABSPV.OUT'      ! EVTS, ACCLIM, IRESTART, OUTFILE
\end{verbatim}
These parameters have the following meaning.

First line: {\tt 46.15D0} - the electron and positron beam energy, {\tt EBEAM}.

Second line: {\tt 24.D-3, 58.D-3, 0.D0} - the electron minimum and maximum
scattering angle (in radians) and the minimum visible energy (in GeV),
{\tt T1MIN, T1MAX, E1MIN}. These cuts
are to be interpreted as ``fiducial'' cuts
within which the events are generated, before going through the triggering
routine.

Third line: the same for the positron, {\tt T2MIN, T2MAX, E2MIN}.

Fourth line: {\tt 0.5D0, 1.D-2, 0.D0, 0.D0, 0.D0} - inputs that may be
required by
the cutting routines for the triggers. These value are stored
in the vector {\tt CALOINPUT(5)} via the common block {\tt COMMON/CALOS}.
In this particular input
the first figure will represent the calorimetric threshold and the second one
the half-opening angle of the electromagnetic jet, while the  other three will
go unused.

Fifth line: {\tt 1} - flag for symmetric cuts, {\tt ISIM}. The user has to
specify if the experimental cuts asked for are ({\tt 1}) or not ({\tt 0})
symmetric  for electron--positron exchange. If they are,
choosing {\tt 1} saves computing time.

Sixth line: {\tt 1} - flag for choosing the triggering routine, {\tt ICALO}.
Possible choices are:
\begin{enumerate}
\item - {\tt TRIGGER1} (sample trigger)
\item - {\tt TRIGGER2} (free slot)
\end{enumerate}

Seventh line: {\tt 1.D4, 0.D0, 0, 'SABSPV.OUT'} -
these are inputs related to the
Monte Carlo integration and to the management of the output. Namely, the total
number of events, {\tt EVTS}, the relative accuracy limit aimed at (the program
stops when the statistical error reaches this level), {\tt
ACCLIM}, the restarting flag, {\tt IRESTART} (if {\tt 1} the program tries to
restart execution from the indicated output file, if {\tt 0} it reinitializes
it), and the output file name, {\tt OUTFILE}.

\section{Test Run Output}
\vskip 10pt

The input file given in Section~\ref{input} produces the following output:
{\small
\begin{verbatim}
 CM ENERGY      =   92.30000

 THETA_EL_MIN   =     .02400
 THETA_EL_MAX   =     .05800
 THETA_POS_MIN  =     .02400
 THETA_POS_MAX  =     .05800
 EN_EL_MIN      =     .00051
 EN_POS_MIN     =     .00051
 CALO INPUTS    =     .50000     .01000     .00000     .00000     .00000

 SIMM. FLAG     =          1

 TRIGGER CHOICE =    TRIGGER1

 CUTOFF         =     .0004758500


 FULL (LL)             =    174.34870 +/-   .35262
 O(ALPHA) (GTGT)       =     -8.85671 +/-   .06548
 BORN PART (GTGT)      =    182.98221 +/-   .29020
 H.O. GTGT AND LL 2,8  =       .22320 +/-   .21581
 VIRTUAL + SOFT (GTGT) =    -91.12947 +/-   .18056
 HARD PHOTON (GTGT)    =    263.66676 +/-   .65744
 ORDER ALPHA (GTGT)    =    172.53729 +/-   .70125
 COMPLETE X-SECT       =    172.76049 +/-   .76344
 COMPL. X-SECT (FACT)  =    172.83542 +/-  1.07296

 NUMBER OF EVENTS:     100000
 RANLUX INITIAL SEQUENCE:      4              1              0              0
 RANLUX RESTARTING SEQUENCE:   4              1        8104045              0

 SUM VECTOR:
         .1743487004930795E+08
        -.8856709767153972E+06
         .1829822083556865E+08
         .2232019045425988E+05
        -.9112947344238646E+07
         .2636667600328065E+08
         .1727604884949618E+08
         .1725372865904168E+08

 SUM^2 VECTOR:
         .4283122616525836E+10
         .5072615600831745E+08
         .4190405492369947E+10
         .4657244067723529E+09
         .1156479410248304E+10
         .1127428205612726E+11
         .8812978799192981E+10
         .7894492407254981E+10
\end{verbatim}
}


\begin{thebibliography}{99}

\bibitem{bhlumi} {S.~Jadach, E.~Ricther-Was, B.F.L.~Ward and Z.~Was, Comput.
Phys. Commun. {\bf 70} (1992) 305; Phys. Lett. {\bf B260} (1991) 438;
Phys. Lett. {\bf B268} (1991) 253.  }

\bibitem{bhagen} {M. Caffo, H. Czy\'z and E.~Remiddi, Phys. Lett. {\bf B327 }
(1994) 369; Int. J. Mod. Phys. {\bf 4} (1993) 591; Nuovo Cimento {\bf 105A}
(1992) 277. }

\bibitem{yr} M.~Cacciari, G.~Montagna, O.~Nicrosini and F.~Piccinini,
in ``Reports of
the Working Group on Precision Calculations for the $Z$ Resonance'', eds.
D.~Bardin, W.~Hollik and G.~Passarino (CERN  Report 95-03, Geneva, 1995),
p.~389, and references therein.

\bibitem{prd}{G.~Montagna, O.~Nicrosini and F.~Piccinini,
Phys.~Rev.~{\bf D48} (1993) 1021.}

\bibitem{noi}{M.~Greco, G.~Montagna, O.~Nicrosini and F.~Piccinini,
Phys.~Lett.~{\bf B318} (1993) 635; \\
G.~Montagna, O.~Nicrosini and F.~Piccinini,
 Comput.~Phys.~Commun. {\bf 78} (1993) 155; erratum, ibid.~{\bf 79}
(1994) 351. }

\bibitem{cr} {M.~Caffo and E.~Remiddi, {\it Bhabha Scattering},
CERN Report 89--08 Vol.~1 (1989) 171, G.~Altarelli, R.~Kleiss and
C.~Verzegnassi Eds., Geneva 1989, and references therein. }

\bibitem{vacpol} {H. Burkhardt, F. Jegerlehner, G. Penso and C. Verzegnassi,
Z.~Phys. {\bf C43} (1989) 497. }


\end{thebibliography}
\end{document}